\def\etal{et~al.}

\def\cm2{cm$^{-2}$}

\def\c2{C~{\sc ii}}
\def\c4{C~{\sc iv}}
\def\fe2{Fe~{\sc ii}}
\def\fe3{Fe~{\sc iii}}
\def\mg1{Mg~{\sc i}}
\def\mg2{Mg~{\sc ii}}
\def\si2{Si~{\sc ii}}
\def\si4{Si~{\sc iv}}
\def\al2{Al~{\sc ii}}
\def\al3{Al~{\sc iii}}
\def\o1{O~{\sc i}}
\def\n1{N~{\sc i}}
\def\h1{H~{\sc i}}

\def\approxlt{\mathrel{\spose{\lower 3pt\hbox{$\sim$}}
        \raise 2.0pt\hbox{$<$}}}
\def\approxgt{\mathrel{\spose{\lower 3pt\hbox{$\sim$}}
        \raise 2.0pt\hbox{$>$}}}

\documentclass[article]{gwp80}
\usepackage{graphicx}
\usepackage{amssymb}
\tabletypesize{\normalsize}  

\def\plotone#1{\centering \leavevmode
\includegraphics[width=.95\columnwidth]{#1}}

\def\plotone#1{\centering \leavevmode
\includegraphics[width=.95\columnwidth]{#1}}

\shortauthors{J.Cohen for the 0Z Project}
\shorttitle{Update -- the 0Z Survey}

\begin{document}
\large    
\pagenumbering{arabic}
\setcounter{page}{101}

\title{A Status Report on the 0Z Project}

%
%
\author{{\noindent Judith G.~Cohen{$^{\rm 1}$}, Norbert Christlieb{$^{\rm 2}$},
Andrew McWilliam{$^{\rm 3}$},
Stephen Shectman{$^{\rm 3}$} \& Ian Thompson{$^{\rm 3}$} \\
\\
{\it 
(1) Palomar Observatory, California Institute of Technology, 
Pasadena, CA, USA\\
(2) Zentrum fur Astronomie der Universitt Heidelberg, Germany\\ 
(3) Observatories of the Carnegie Institution of Washington, 
Pasadena, CA, USA\\} 
}
}
%
%
\email{(1) jlc@astro.caltech.edu (2) N.Christlieb@lsw.uni-heidelberg.de
(3) andy@obs.carnegiescience.edu,\ 
shec@obs.carnegiescience.edu,\  
ian@obs.carnegiescience.edu }

\begin{abstract}
We present an update on the status of the 0Z Survey, an effort
to datamine the Hamburg/ESO Survey to find and study in detail
a large sample of extremely metal-poor
Galactic halo field stars with [Fe/H] $< -3.0$~dex.
After searching 1,565 moderate resolution spectra of candidates
selected from the HES, we have acquired high resolution
spectra of 103 of the most metal-poor of them.  Detailed
abundance analyses have been performed for all these stars.  
This has resulted in the discovery of 18 new stars below
[Fe/H] $-3.5$~dex and 57 below $-3.0$~dex. 
Some  results are presented regarding our search for outliers
in chemical abundances of particular species among
the sample of 103 stars.  Ignoring C and N, about 15\%
of the sample are ``abnormal'' in some way.  Our plans to
complete this project and write the final set of papers
are described.

\end{abstract}

\section{Introduction}

Extremely metal poor stars were presumably among the first
stars formed in the Galaxy, and hence represent in effect
a local high-redshift population.  It is from this paradigm that the
term ``galactic archeology'' arises.  Such stars
provide important clues to the chemical
history of our Galaxy, the role and type of early SN, the
mode of star formation in the proto-Milky Way, and the formation
of the Galactic halo.  The number of extremely metal poor (EMP)
stars known as of 2005 is summarized by Beers \& Christlieb(2005);
it has grown slowly since that time, but the sample is
still quite small. Our goal is to increase this sample
substantially so that statistical studies 
of the Galactic EMP star population become feasible.

\section{Sample Selection Strategy} 

We are engaged in a long term program
to datamine the Hamburg/ESO Survey for extremely metal-poor stars
with [Fe/H] $< -3$~dex, i.e. less than 1/1000 the Solar
Fe/H ratio\footnote{The 
standard nomenclature is adopted; the abundance of
element $X$ is given by $\epsilon(X) = N(X)/N(H)$ on a scale where
$N(H) = 10^{12}$ H atoms.  Then
[X/H] = log$_{10}$[N(X)/N(H)] $-$ log$_{10}$[N(X)/N(H)]$_{\odot}$, and similarly
for [X/Fe].}.  The HES is an objective prism survey
with 
low spectral dispersion taken with photographic plates
of high-latitude Galactic halo fields.
Given the low spectral resolution and, at the faint end of the sample,
the modest SNR, the best that can be done
is to produce lists of candidate EMP stars.  Since the entire
survey, although taken with photographic plates, was subsequently digitized,
the selection of candidates was carried out using well defined
and reproducible scripts and procedures, which can be modified
based on experience, and the selection 
redone as necessary at later times.

The candidate lists thus generated are unreliable, with many 
contaminating higher metallicity stars.  Thus the EMP candidates
selected from the HES database
must be verified by moderate resolution spectra taken
at 5~m class telescopes.  We observed 1,565 candidates
with the 6.5~m Magellan  Clay Telescope or the Double Spectrograph 
Oke \& Gunn(1982) at the 
5~m Hale Telescope at Palomar Mountain.  Of these
26 were rejected as not Galactic halo field stars.
Most of these interlopers are dMe stars.

By summer 2006, when observations at the P200
ended, followup spectroscopy for candidates in the fall north field 
(approximately 900 sq deg on the sky
north of Dec $-20^{\circ}$ covered by the HES) was 99\% complete
to the limit of the HES at $B = 17.5$~mag.
The spring north field was complete to B = 16.5~mag,
and missing only 5 stars to B = 16.6 mag.

After each such observing run, a metallicity
[Fe/H](HES) was determined from the followup spectrum of each candidate
observed.  This was based on the strength of
absorption at H$\delta$, a measure
of stellar temperature, coupled with that of
the 3933~\AA\ line of Ca~II, which provided
the metallicity given $T_{eff}$.  The
algorithm of Beers et al.~(1999)
was used for this purpose.

Since this 
algorithm was improved with time, especially with
the discovery by Cohen et al.~(2005) of its problems in
accurately predicting
[Fe/H] from spectra of highly C-enhanced stars, 
all the moderate
resolution spectra were re-processed at the end
of this campaign in early 2007.  At that time {\it{no}}
[Fe/H](HES) value was assigned to highly C-rich stars.  
Furthermore the selection function from the 
HES evolved somewhat
with time, and some stars which were on the original
target list for followup spectroscopy were subsequently deleted, while
others were added.

The accuracy of the values of
[Fe/H](HES) as derived from the moderate resolution
spectra was tested with multiple observations of
a set of well studied low metallicity stars
from the HK Survey Beers, Preston \& Shectman(1985, 1992)
as well as multiple
observations of the candidates themselves.

These [Fe/H](HES) comprised a major part of the statistically
complete sample used by Schorck et al.~(2009) to determine
the metallicity distribution function in the Galactic halo, with
emphasis on the EMP tail.  This large sample, combined
with careful consideration of selection
effects, led to important new conclusions about the similarity
of the MDF 
of the Milky Way halo MDF and that of its dSph satellites
in the extremely metal-poor regime.

\section{High Resolution Observations and Data Reduction}

Since our goal is a detailed characterization of a statistically
significant population of confirmed EMP field halo stars,
we attempted to get high spectral resolution observations
with either MIKE on the 6.5~m Magellan  Telescope at
Las Campanas or HIRES Vogt et al.~(1994) at the 10~m Keck~1 Telescope for
all stars which were assigned [Fe/H](HES) $< -3$~dex.
Stars at slightly greater [Fe/H](HES), which were of course
more numerous, were observed to fill in scheduled nights
with MIKE/Magellan or HIRES/Keck
when necessary.  
The earliest high dispersion observing
runs (i.e. those in 2001-2002) included some EMP candidates
suggested by N.~Christlieb or T.~Beers, as these runs 
occurred before the 0Z Survey
could build up a substantial set of verified EMP candidates
from its own followup spectra. The
sample presented here includes 99 stars observed
with HIRES and four with MIKE, for a total sample
of 103 stars.  

\begin{figure*}
\centering
\plotone{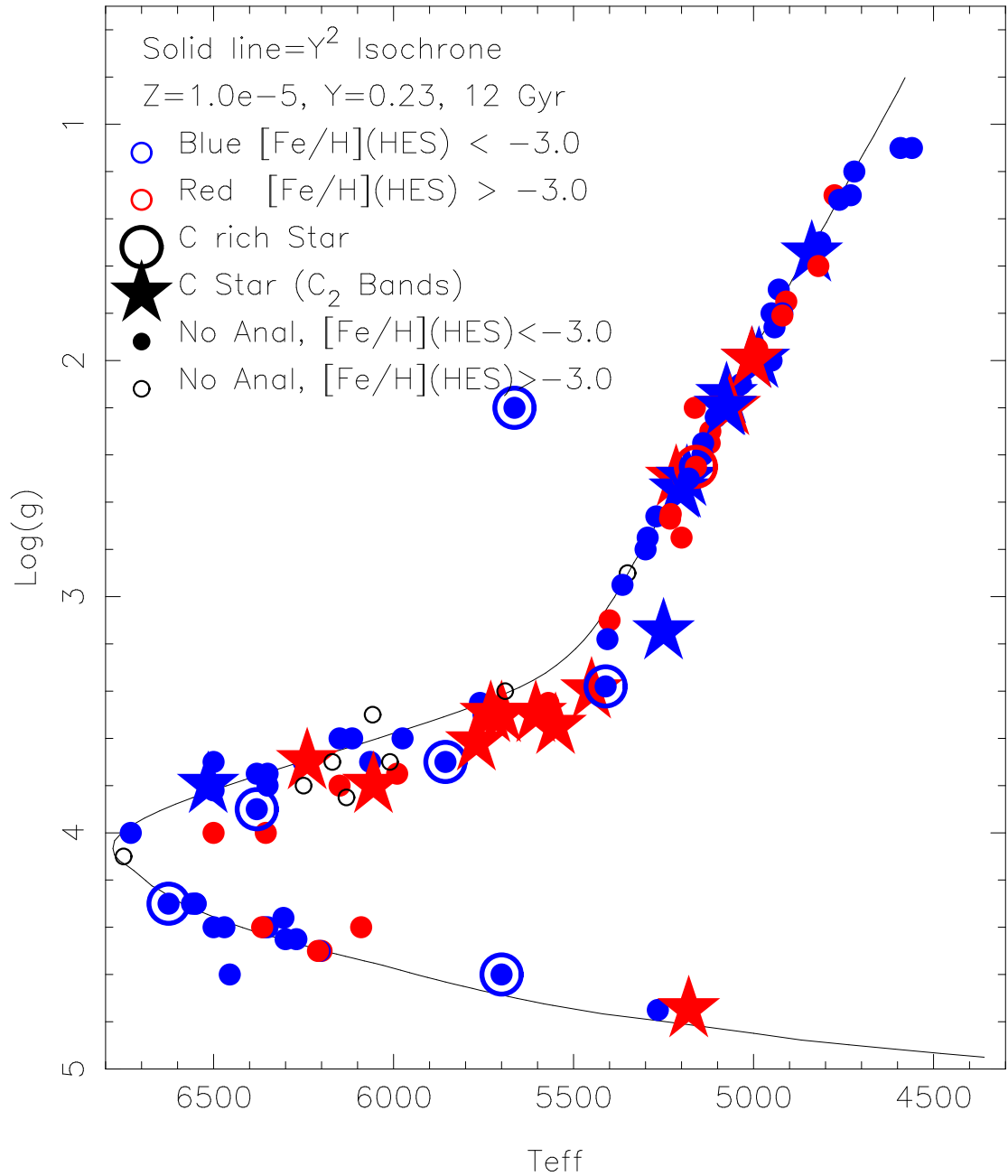}
\vskip0pt
\caption{The sample of stars with high dispersion observations is
shown on the $T_{eff}$  -- log$(g)$ plane together with a low
metallicity Y$^2$ 12 Gyr isochrone.  Those stars
with [Fe/H](HES) $< -3.0$~dex are indicated, as are C-rich stars.
\label{figure_cmd} }
\end{figure*}

All of these spectra were analyzed by J.~Cohen 
in a homogeneous manner with identical procedures,
stellar parameter determinations, line lists,
transition probabilities, etc with the help
of post-doctoral fellows Jorge Melendez and Wenjin Huang.
We used the LTE spectrum synthesis program MOOG (Sneden 1973) and the grid
of plane-parallel (i.e. 1D) LTE model atmospheres from 
Kurucz (1993) to compute abundances
from the measured equivalent widths.
The effective temperatures were derived from broad-band
optical and 2-MASS infrared photometry.  Our program
using the
Andicam queue at CTIO was the primary source for
$V,~I$ optical photometry.  Results from SDSS 
were used when available, and
as a last resort V(HES) was used for a very small
number of stars.  Our analysis uses the nominal
$T_{eff}$ from $V-I$, $V-J$, and $V-K_s$.  Surface
gravity is defined by assuming that the star lies
along a very metal-poor Y2 isochrone.  Because
of uncertainties in the photometry, we felt free to
slightly adjust $T_{eff}$ if necessary to achieve
a good result.  Although the nominal assumption was
that all stars were more luminous than the main
sequence turnoff, occasionally we were forced to 
adjust log($g$) to set a
star below the turnoff.  Fig.~\ref{figure_cmd}
shows the location of the sample in the $T_{eff}$ -- log($g$) plane.
Candidates which turned out to be EMP stars 
are indicated, as are those that have highly enhanced
carbon abundance.

\begin{figure*}
\centering
\plotone{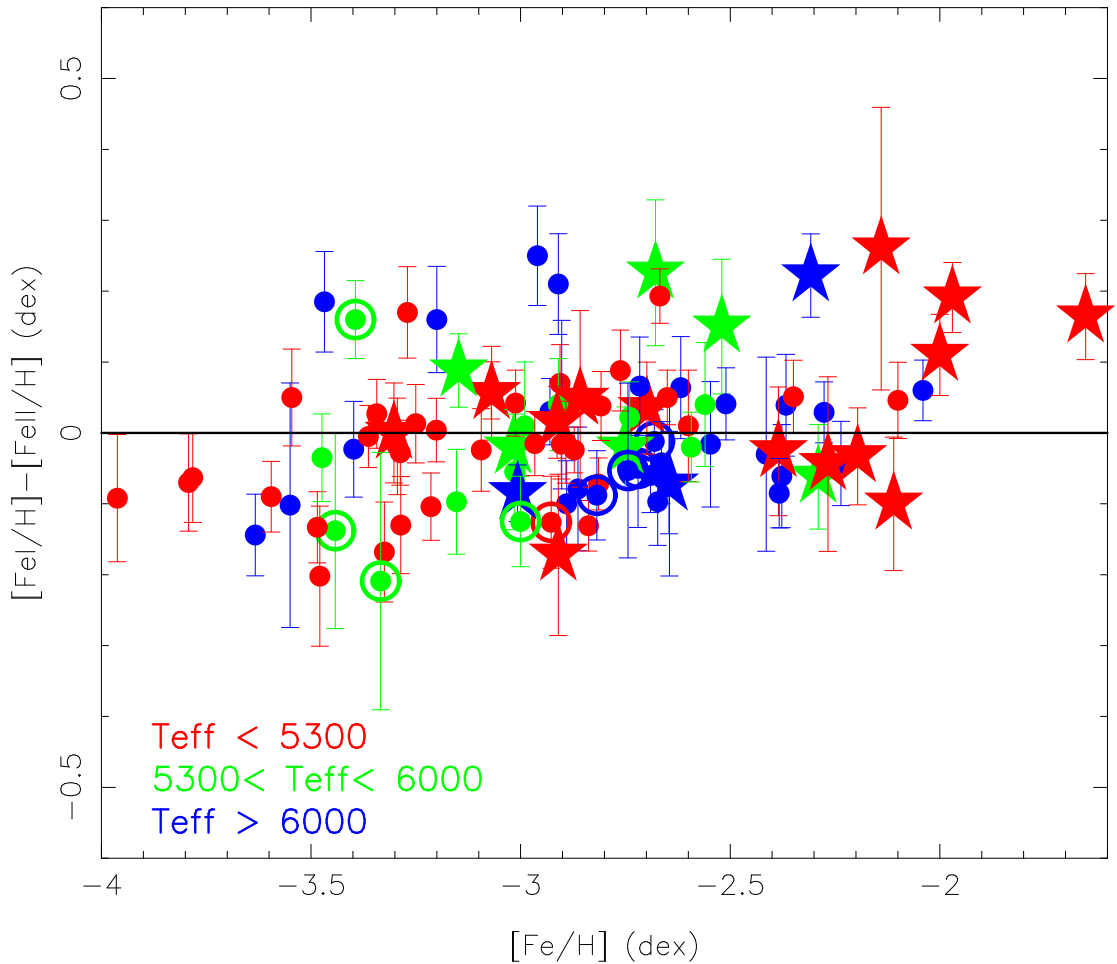}
\vskip0pt
\caption{The ionization equilibrium for Fe~I vs Fe~II is shown
as a function of [Fe/H] (from Fe~I) for the HES EMP candidates with
detailed abundance analyses.  
Red points
have $T_{eff} <  5300$~K, green $5300 < T_{eff} < 6000$~K,
blue $T_{eff} > 6000$~K.  Stars denote carbon stars,; circled
points are highly C-rich, but do not show C$_2$ bands.
\label{figure_feion} }
\end{figure*}

One test of the validity of the abundance analyses
is to examine the ionization equilibrium of Fe,
i.e. the deduced [Fe/H] from lines of Fe~I versus
those of Fe~II, for each star.  This is shown in
Fig.~\ref{figure_feion}, and looks quite reasonable.
The mean difference between the Fe abundance derived
from the neutral and the singly ionized species 
for 100 EMP candidates from out sample with
with HIRES spectra and detailed abundance
analyses by J.~Cohen
is
0.00~dex with $\sigma ~ = ~ 0.10$~dex 

\section{Current Status of the 0Z Project}

Detailed abundance analyses by J.~Cohen give [Fe/H] values
systematically slightly higher than those which would be derived
by most other large surveys in this field.  The
reasons for this small offset of $\sim$0.15~dex in [Fe/H] are understood,
and are discussed in an appendix in Cohen et al.~(2008).   In the [Fe/H] scale
of the VLT First Stars Survey 
(Cayrel et al.~2004),
we have found 18 new stars with
[Fe/H] below
$-3.5$~dex  and 57 new stars below $-3.0$~dex.
These represent a substantial increase in the number
of EMP stars known.

\begin{figure*}
\centering
\plotone{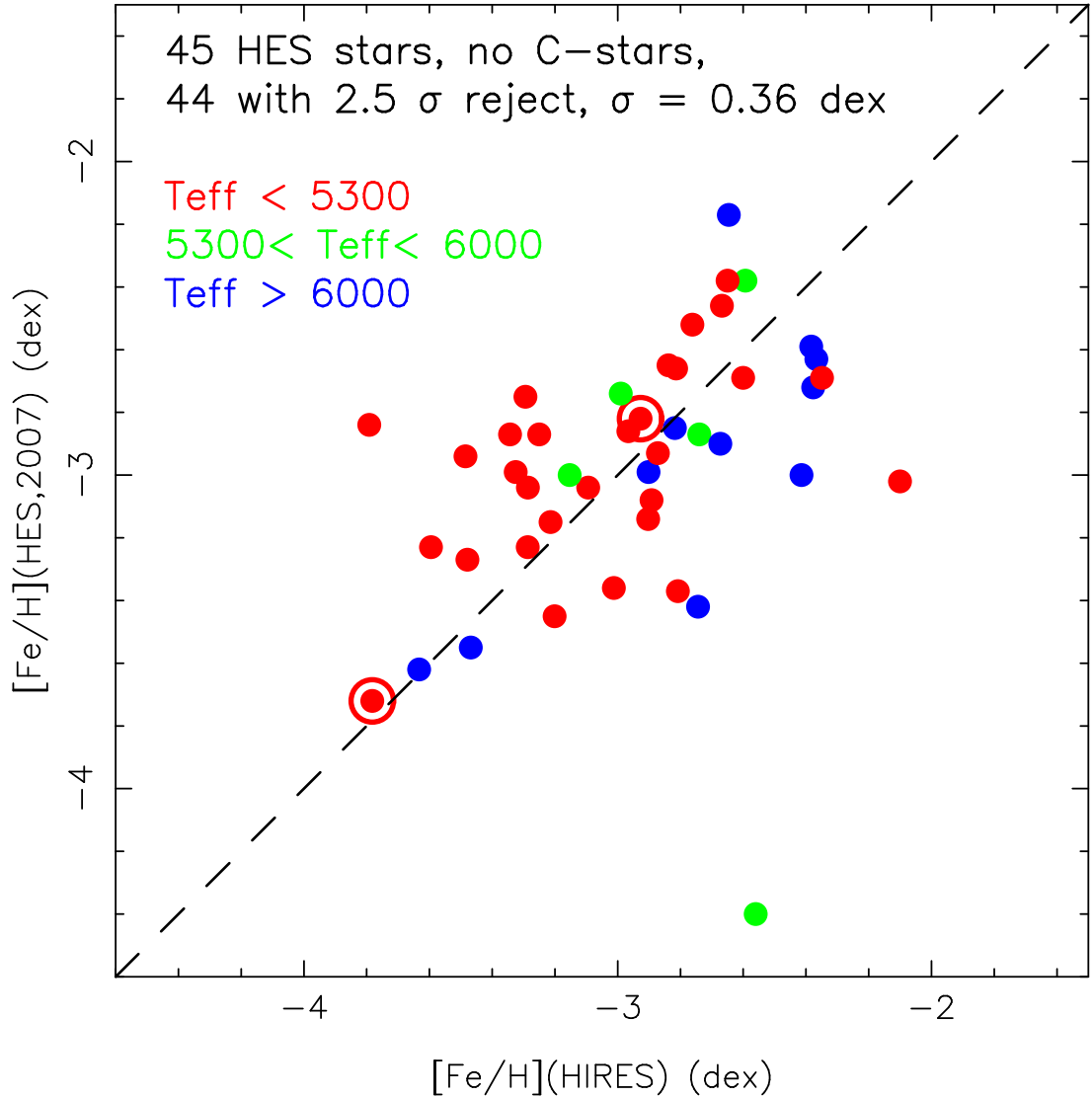}
\vskip0pt
\caption{A comparison of [Fe/H](HES) from the moderate resolution
followup spectra to [Fe/H] derived from the high dispersion spectra
with a detailed abundance analysis. Highly C-rich stars
are ignored. Colors denote the three
ranges of $T_{eff}$ as in Fig.~\ref{figure_feion}.
Only a small systematic trend with $T_{eff}$ is seen.
\label{figure_hes_hires} }
\end{figure*}

Ignoring the C-rich stars, a comparison of
the derived [Fe/H](HIRES) with that from the
moderate resolution spectra is shown in Fig.~\ref{figure_hes_hires}.
Excluding one very discrepant star, the $\sigma$
of the differences is 0.36~dex, most of which presumably
should be ascribed to the uncertainty in [Fe/H](HES).

\begin{figure*}
\centering
\plotone{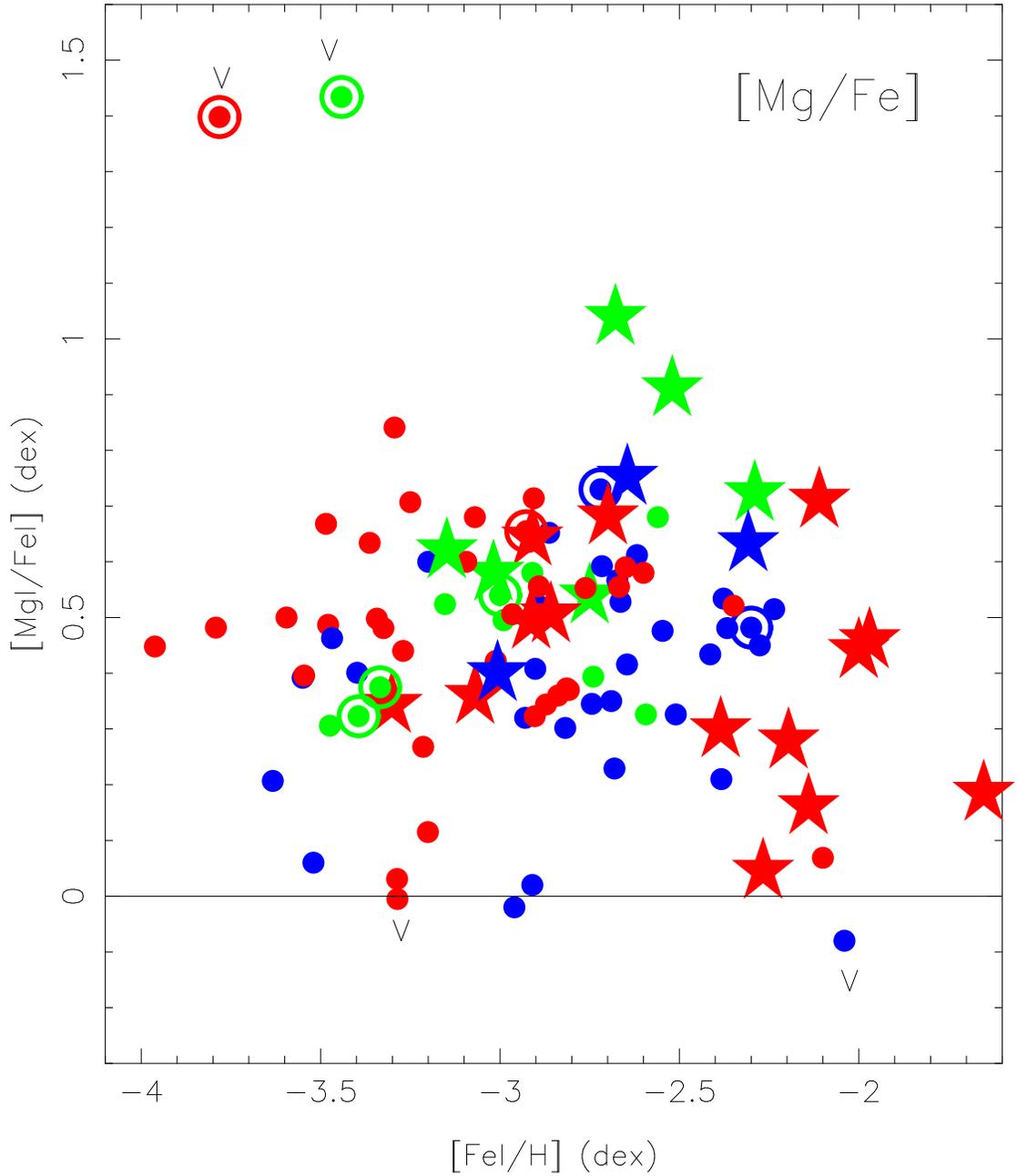}
\vskip0pt
\caption{[Mg/Fe] vs [Fe/H] for the HIRES sample.
Colors denote the three
ranges of $T_{eff}$ as in Fig.~\ref{figure_feion}.
Stars denote carbon stars; circled
points are highly C-rich, but do not show C$_2$ bands. ``V'' denotes
abundance ratios that have been carefully checked and appear correct: 
those that may not be valid when carefully checked are marked 
with ``?''.
\label{figure_hes_mgfe}}
\end{figure*}

\begin{figure*}
\centering
\plotone{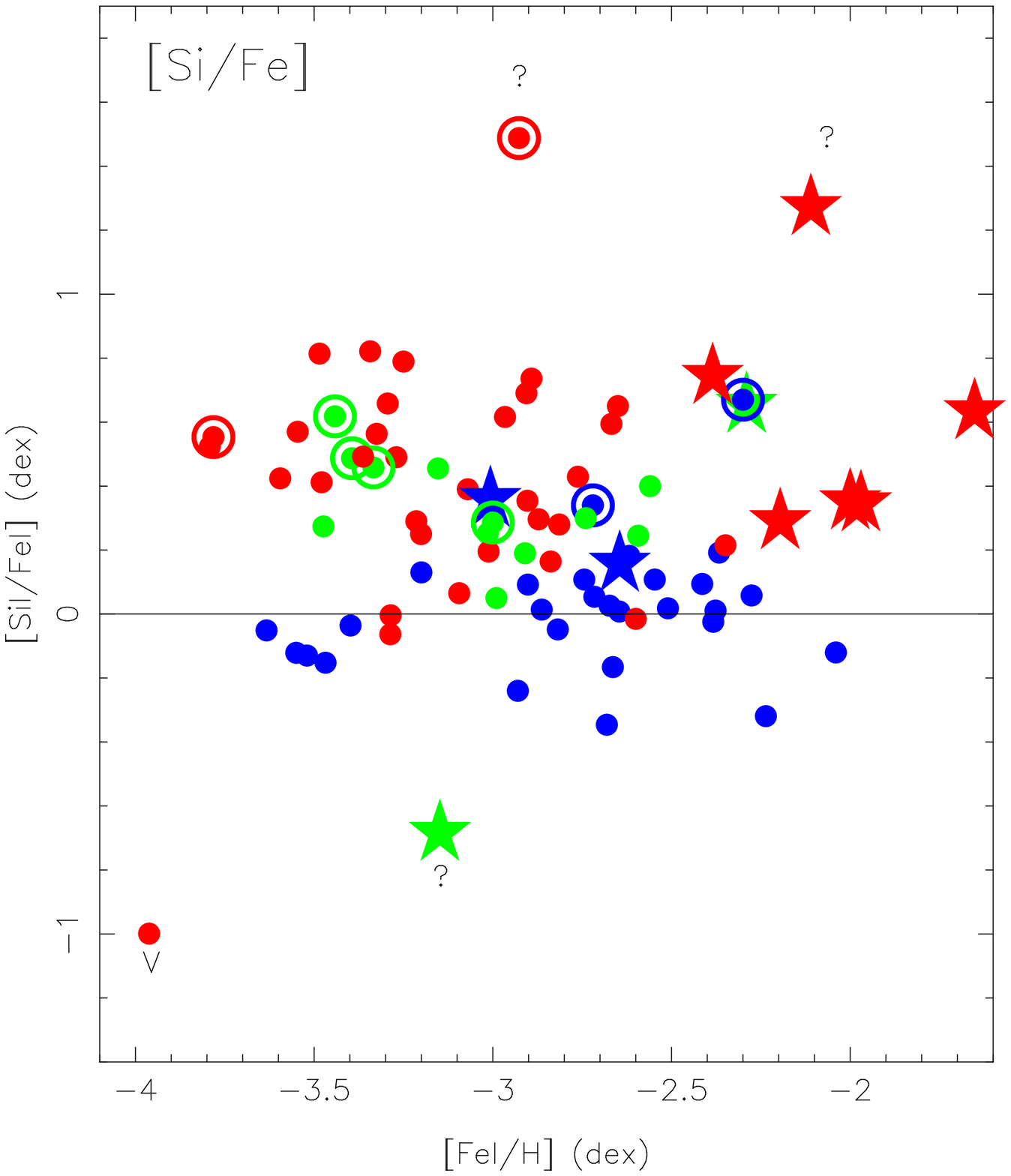}
\vskip0pt
\caption{The same as Fig.~\ref{figure_hes_mgfe} for [Si/Fe].
The symbols are the same.  In this case there is a trend of
[Si/Fe] at a fixed [Fe/H] with $T_{eff}$, suggesting a problem
in the analysis.
\label{figure_hes_sife} }
\end{figure*}

Once this sample of 103 homogeneous abundance
analyses of supposed EMP stars was assembled,
we constructed diagrams of relative abundance trends versus
overall metallicity, i.e. of [X/Fe] versus
[Fe/H].  As examples, we show
the trends for [Mg/Fe] and for
[Si/Fe]  in Fig.~\ref{figure_hes_mgfe} and
\ref{figure_hes_sife}.  For each such plot, we have begun to check each of
the outlier stars, looking for errors, examining
the original spectra at the wavelengths of crucial lines
when necessary, etc.  So far between 4 and 8 outliers
for each of 10 elements have been checked,
with perhaps 10 more elements left to do.  Those stars
for which we have carried out such a check
for a particular species and did not find any
problem nor any reason not to believe the result
are indicated by ``V'' in the plots, while those
for which we could not be sure the result was valid
are marked with ``?''.  The latter usually
happened for C-rich stars with strong 
molecular bands and with spectra taken early in the project prior to 
the HIRES detector upgrade in 2004.  These early
spectra do not extend far enough into the red
to reach some key lines beyond the CH, CN, 
or the 5170~\AA\ C$_2$ bands.  Hence the abnormal
relative abundance ratios initially derived
may not be valid.  Note that the two very high [Mg/Fe] values
for two C-rich stars with [Fe/H] $< -3.5$~dex seen in the upper
left corner of Fig.~\ref{figure_hes_mgfe} have been carefully
checked and appear valid.  Large enhancements of Mg have been
reported in a few cases among very metal-poor C-rich stars
(e.g., see McWilliam et al 1995, Aoki et al. 2002).

In the case of Fig.~\ref{figure_hes_sife} the
dependence of [Si/Fe] on $T_{eff}$ noted
earlier by 
Preston et al.~(2006) (see their Fig.~10)
is apparent in our abundance analyses as well.
Presumably the problem lies in non-LTE corrections
which vary significantly with $T_{eff}$.  The most metal-poor
star in this figure has a very low [Si/Fe]; it is HE~1424--0241,  the very
peculiar star whose discovery was announced in
Cohen et al.~(2007) and was discussed in detail in Cohen et al.~(2008).

Among the sample of 103 stars, we found one
new extreme $r$-process enhanced star
with [Eu/Fe] +1.5~dex.  It is somewhat hotter
than the prototype of this class, CS22892--052, and
is very slightly higher in [Fe/H].

\section{Ongoing Work to Finish the 0Z Project}

Although the 0Z Project has published
a large number of papers 
on specific interesting stars found by the 0Z project
(see, e.g. Lucatello et al 2003, Cohen et al 2003,
Cohen et al 2004, Cohen et al 2006, Cohen et al 2007,
Cohen et al 2008)
%
%
our current
goal is to finish the project by  writing a set of papers which will present
the entire sample of EMP candidates selected from the HES with
moderate resolution spectra (1,565 stars).  Another paper
will give the
detailed abundance analyses for
the entire set of stars with HIRES spectra (103 stars),
and describe the implications for early nucleosynthesis and
supernovae in the young Milky Way
 that can be derived from this
work.  In particular, with such a large homogeneous
data set we will
be able to  construct
better
comparisons and detect modest outliers from the mean
trends more easily.

After we finish checking all the outliers, we
need to define a better way of visualizing the interlinking
of abnormalities in one specific species with
the results for all other species detected
in stars which are outliers as compared to 
stars which do not show any abnormalities.
We
are doing this by selecting the closest
stars in $T_{eff}$ to the peculiar star
from our large sample as comparison objects
and examining plots of [X/Fe] versus atomic number of
the species X for the suspected outlier versus
the comparison stars.

The process of checking the outliers and linking
abnormalities across all the species observed
is currently underway.  Because of our large
homogeneous sample of EMP stars with 
homogeneous detailed
abundance analyses we can find even modest
outliers.  Ignoring enhanced carbon and/or nitrogen, 
thus far approximately 15 of the 103 stars
are peculiar in some way, but the vast
majority are ``normal'' EMP stars
which to within the uncertainties
obey well defined trends of [X/Fe] 
vs [Fe/H] for
all species observed.

\section{Acknowledgements}

We are very grateful
to the Palomar, Las Campanas, and Keck time allocation committees for
their long-term support of
this  campaign during the initial phase of moderate resolution 
spectroscopy which began
began in 2000 and ended in 2006 as well as the subsequent high resolution
spectroscopy.
J.~Cohen acknowledges partial support from NSF grants
AST--0507219 and AST--0908139.
We are grateful to the many people  
who have worked to make the Keck Telescope and its instruments  
a reality and to operate and maintain the Keck Observatory. 
The authors wish to extend special thanks to those of Hawaiian ancestry
on whose sacred mountain we are privileged to be guests. 
Without their generous hospitality, none of the observations presented
herein would have been possible.

\end{document}